\documentclass{moriond}

\def\Journal#1#2#3#4{{#1} {\bf #2}, #3 (#4)}

\def\NPB{{\em Nucl. Phys.} B}

\def\PRL{\em Phys. Rev. Lett.}
\def\PRD{{\em Phys. Rev.} D}

\def\be{\begin{equation}}
\def\ee{\end{equation}}
\def\bea{\begin{eqnarray}}
\def\eea{\end{eqnarray}}

\usepackage{amsmath}
\usepackage{amssymb}

\newcommand{\Photo}{\includegraphics[height=35mm]{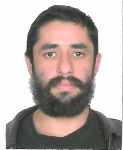}}

\begin{document}
\vspace*{4cm}
\title{A NEW LONG DISTANCE CONTRIBUTION TO $B^\pm\to K^\pm/\pi^\pm\ell^+\ell^-$ DECAYS}

\author{ A. GUEVARA, G. L\'OPEZ CASTRO, P. ROIG, S. L. TOSTADO }

\address{Departamento de Física, Cinvestav del IPN, Apdo Postal 14-740, 07000 Mexico DF. Mexico
}
\maketitle\abstracts{
 We have identified a missing long-distance (one-photon exchange) contribution to the $B^\pm\to K^\pm/\pi\ell^+\ell^-$ decays. Although it does not 
 help to explain the anomaly measured by the LHCb Coll. in the ratio $R_K = \mathcal{B}(B^\pm\to K^\pm\mu^+\mu^-)/\mathcal{B}(B^\pm\to K^\pm e^+e^-)\sim0.75$ 
 (2.6 $\sigma$ deviation with respect to $R_K=1+\mathcal{O}(10^{-4})$ in the Standard Model), it provides a sizable contribution to the 
 branching ratio of the $B^\pm\to \pi^\pm\ell^+\ell^-$ decays. The new decay mechanism gives rise to a measurable CP asymmetry, which is of order 1\% in both channels. 
 These predictions can be tested in forthcoming LHCb measurements. All the details can be seen in\cite{this}.
 }

\section{Introduction}

In the search of a more fundamental description of interactions in nature, one has to look for effects that can not be described by the 
current theory of particle physics, namely the Standard Model (SM). A way to do this is studying the more suppressed processes in the SM, 
where a deviation from a precise SM prediction is expected to be more significant than in processes that are not suppressed. In order 
to claim that an effect comes from physics Beyond the SM (BSM) in a certain process, all the SM contribution have to be known and understood 
at the precision required by experiments. 
In $B^\pm\to K^\pm\ell^+\ell^-$ decays, LHCb\cite{RK} has measured the $R_K$ ratio in the $[1,6]$ GeV$^2$ region of squared lepton pair
invariant mass $q^2$, which is $R_K=0.745^{+0.090}_{-0.074}\pm0.036$, 
while the current SM prediction in the same energy range is\cite{SDK} $R_K^{SM}=1+(3.0^{+1.0}_{-0.7})\times10^{-4}$, which takes into account 
only Short Distance (SD) contributions to the process. This LHCb measurement, together with a previous prediction\cite{us} of lepton universality 
violation in another process due to purely kinematic effects, prompted us to study Long Distance (LD) contributions to the Branching Fraction of 
$B^\pm\to P^\pm\ell^+\ell^-$ decays, with $P=\pi,K$.
 
   \begin{figure}
    \centering
    \includegraphics[scale=.65]{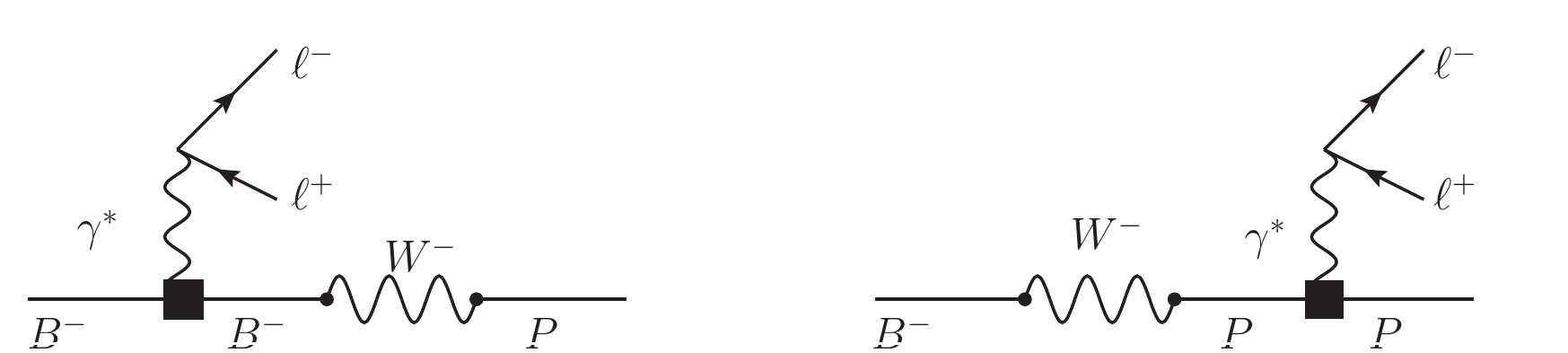}\caption{New long distance contribution to the process, where the square is for structure dependent one-photon exchange. 
    Other structure dependent contributions vanish due to gauge invariance.
    }\label{fotonX}
   \end{figure}

 \section{Short Distance contribution}
 
   In order to compute the SD amplitude to the $B^-\to K^-/\pi^-\ell^+\ell^-$ decay, we follow previous computations of such amplitude\cite{SDK,Khod} using 
   QCD Factorization (QCDF), where the 
   effective weak hamiltonian 
 \begin{equation} 
  \mathcal{H}_{eff}=-\frac{G_F\alpha}{\sqrt{2}\pi}V_{q'b}V_{q'q}^*\sum_i C^{q}_i(\mu_s) O^{q}_i(\mu_s)
 \end{equation}
   is used to compute this amplitude, where $q'=c,t$ and $q=d,s$. The main contributions to this process come from the operators with 
   the vector and axial lepton current $O^q_9=(\bar{q}\gamma_\mu b_L)(\bar{\ell}\gamma^\mu\ell)$ and $O^q_{10}=(\bar{q}\gamma_\mu b_L)(\bar{\ell}\gamma^\mu\gamma_5\ell)$, 
   nevertheless at higher order in QCDF all the other operators will contribute, but less significantly than the former operators.

   The leading contribution to the amplitude computed from this hamiltonian is the following
   
   \begin{equation}
    \mathcal{M}[B^-\to P^- \ell^+\ell^-]=\frac{G_F\alpha}{\sqrt{2}\pi}V_{q'b}V_{q'q}^*\xi_P(q^2)p_B^\mu\left(F_V \bar{\ell}\gamma_\mu\ell
  + F_A \bar{\ell}\gamma_\mu\gamma_5\ell\right),
   \end{equation}
where $\xi_P$ is the soft form factor associated to the $B\to P$ transition, and the $F_V$ and $F_A$ are form factors that depend mainly on SD 
effects, which becomes evident when their dependence on the so called Wilson Coefficients ($C_1,...,C_{10}$) is noticed. The $q^2$ dependence 
of the soft form factor can be obtained, for example, using Light Cone Sum Rules (LCSR) (for more details see\cite{Ball}), 
   \begin{subequations}
    \begin{align}
     \xi_\pi(q^2)&=\frac{0.918}{1-q^2/(5.32\text{ GeV})^2} - \frac{0.675}{1-q^2/(6.18\text{ GeV})^2} 
     + \mathcal{P}_\pi(q^2)\\
     \xi_K(q^2)&=\frac{0.0541}{1 - q^2/(5.41\text{ GeV})^2} + \frac{0.2166}{\left[1 - q^2/(5.41\text{ GeV})^2\right]^2}
     + \mathcal{P}_K(q^2)     
    \end{align}
   \end{subequations}
where it is fitted to dipole expressions, which are the main contributions to this form factor. 
   The $\mathcal{P}_P(q^2)$ are polynomials on $q^2$, whose coefficients correspond to the Gegenbauer moments. The values of $C_9$ and $C_{10}$ 
   are taken from the NNLL calculation\cite{Ben}.

   \begin{figure}
    \centering
    \includegraphics[scale=.28]{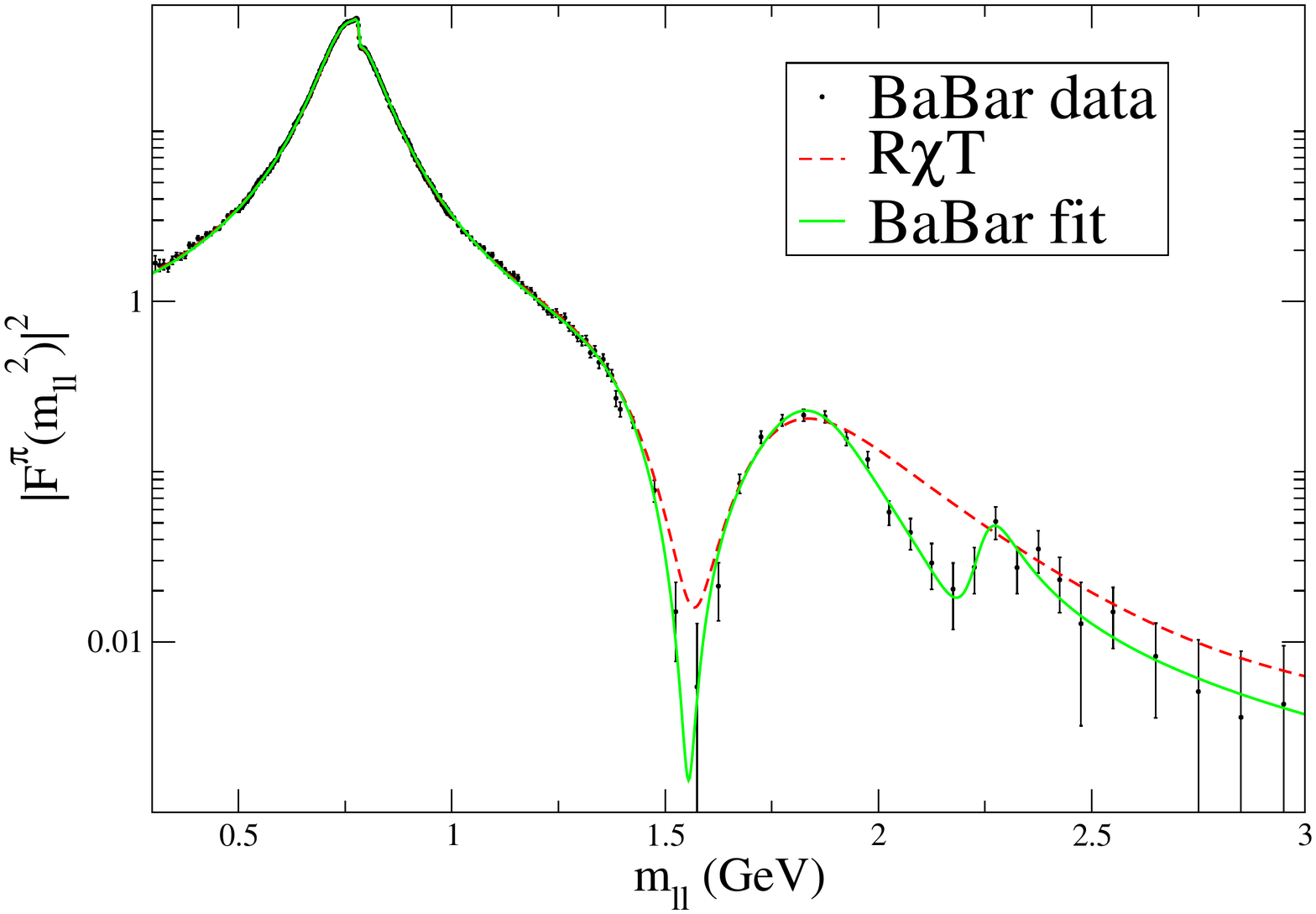}
    \includegraphics[scale=.28]{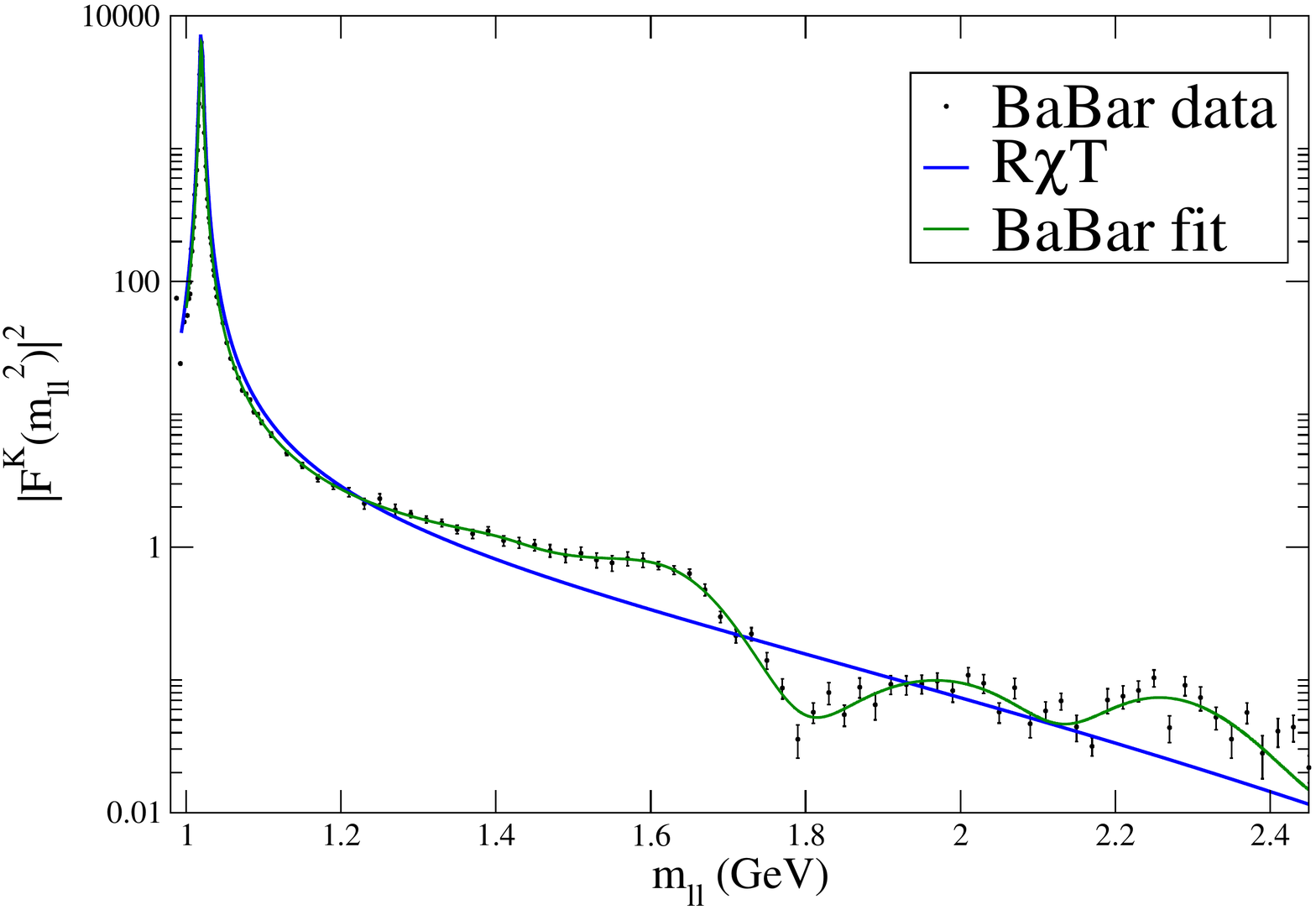}
    \caption{Our form factor compared to data and BaBar fitted form factor for $P=\pi$ (left) and $P=K$ (right). The form factors depend on 
    the dilepton pair invariant mass.}\label{FF}
   \end{figure}

   \section{New Long Distance contribution}

   We focus on the same energy range LHCb uses, which excludes completely any charm quark contamination.
   The contributing Feynman diagrams are those of Fig.\ref{fotonX}; other contributions to this process from purely LD effects 
   vanish due to electric charge conservation\cite{Ecker}. The left-hand-side diagram on Fig.\ref{fotonX} is suppressed by a factor $m_P^2/m_B^2$ with respect to the 
   right-hand-side diagram, so that we can neglect its contribution. The LD contributions for the process are obtained using Resonance Chiral 
   Theory (R$\chi$T)\cite{Chi}.

   The amplitude that we obtain from the leading contribution is 
   
   \begin{equation}
    \mathcal{M}_{LD}=\sqrt{2} G_F V_{ub}V_{uq}^*f_Bf_P\frac{e^2}{q^2}\frac{m_B^2}{m_B^2 - m_P^2}[F_P(q^2) - 1]p_B^\mu\bar{\ell}\gamma_\mu\ell,
   \end{equation}
which we see that has the same structure as the $F_V$ part of the SD amplitude, so that it can be added as a correction to this form factor in the following way
   
   \begin{equation}
    F_V^{eff}= F_V + \frac{\kappa_P m_B^2}{q^2}\frac{F_P(q^2)-1}{\xi_P(q^2)},
   \end{equation}
where the $\kappa_P=-8\pi^{2}\frac{V_{ub}V^*_{uq}}{V_{tb}V^*_{tq}}\frac{f_B f_P}{(m_B^2 - m_P^2)}$ is a dimensionless constant $\mathcal{O}(10^{-2})$ 
   times a CKM suppression factor which depends on the final $P$ state. In Wolfenstein's parametrization, for $P=\pi$, $\kappa$ is $\mathcal{O}(\lambda^0)$, 
   while for $P=K$ is $\mathcal{O}(\lambda^{2})$.\\
   
   Our ignorance of the underlying dynamics is encoded in the $F_P(q^2)$ form factors, which can be seen in Fig.\ref{FF}. These are almost saturated by 
   the lowest-lying light-flavor resonances ($\rho$, $\omega$ and $\phi$), following different parametrizations for\cite{Olga} $P=\pi$ and\cite{Arga} $P=K$.
   We have also used phenomenological models by BaBar\cite{BaB} as a test of our error. Their form factors are fitted using also  
   heavier resonances. The almost perfect agreement on the peaks of the lightest resonances ensures a prediction with small error because the contribution of
   heavier resonances is negligible.

   \begin{figure}
    \centering
    \includegraphics[scale=.28]{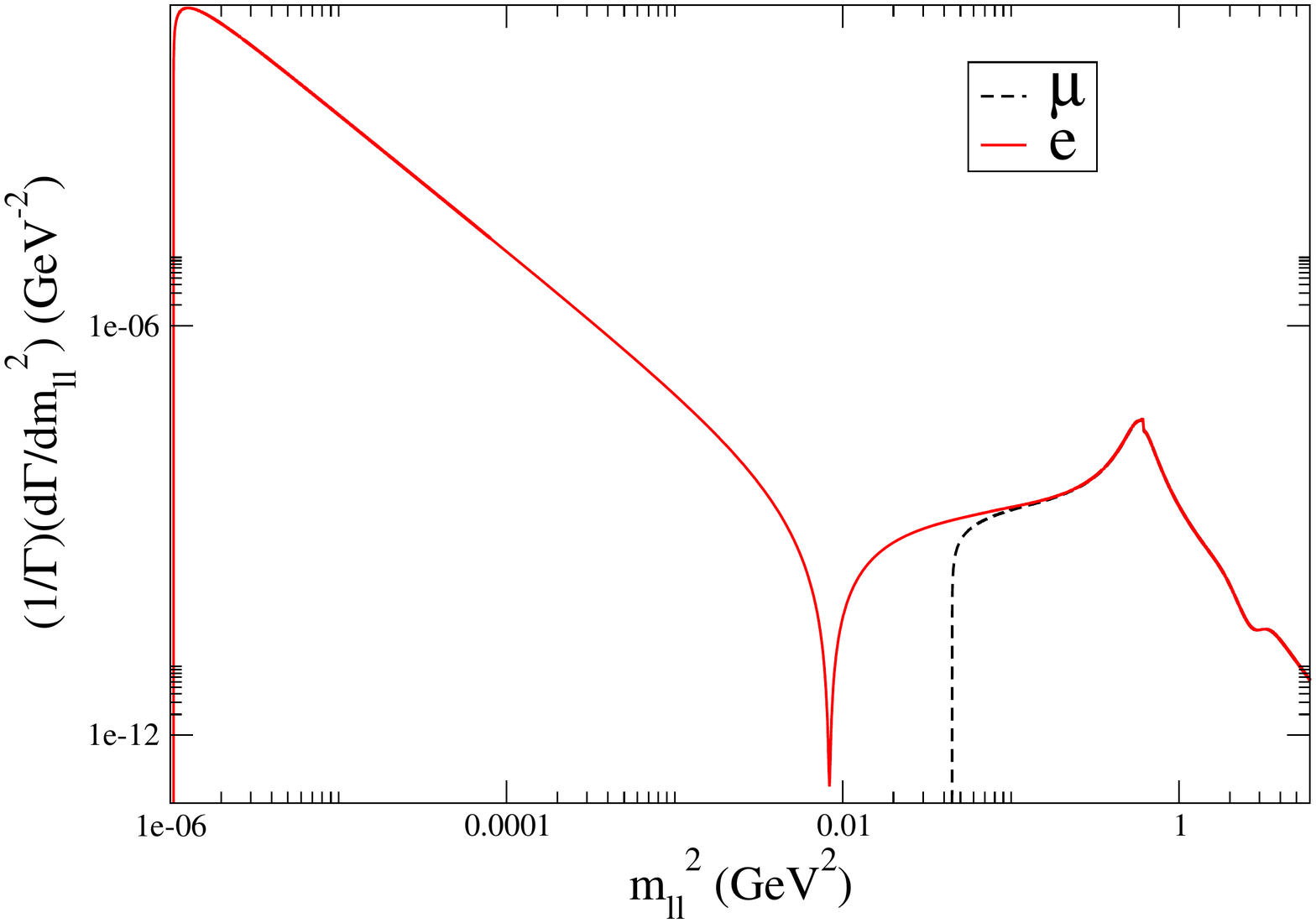}
    \includegraphics[scale=.28]{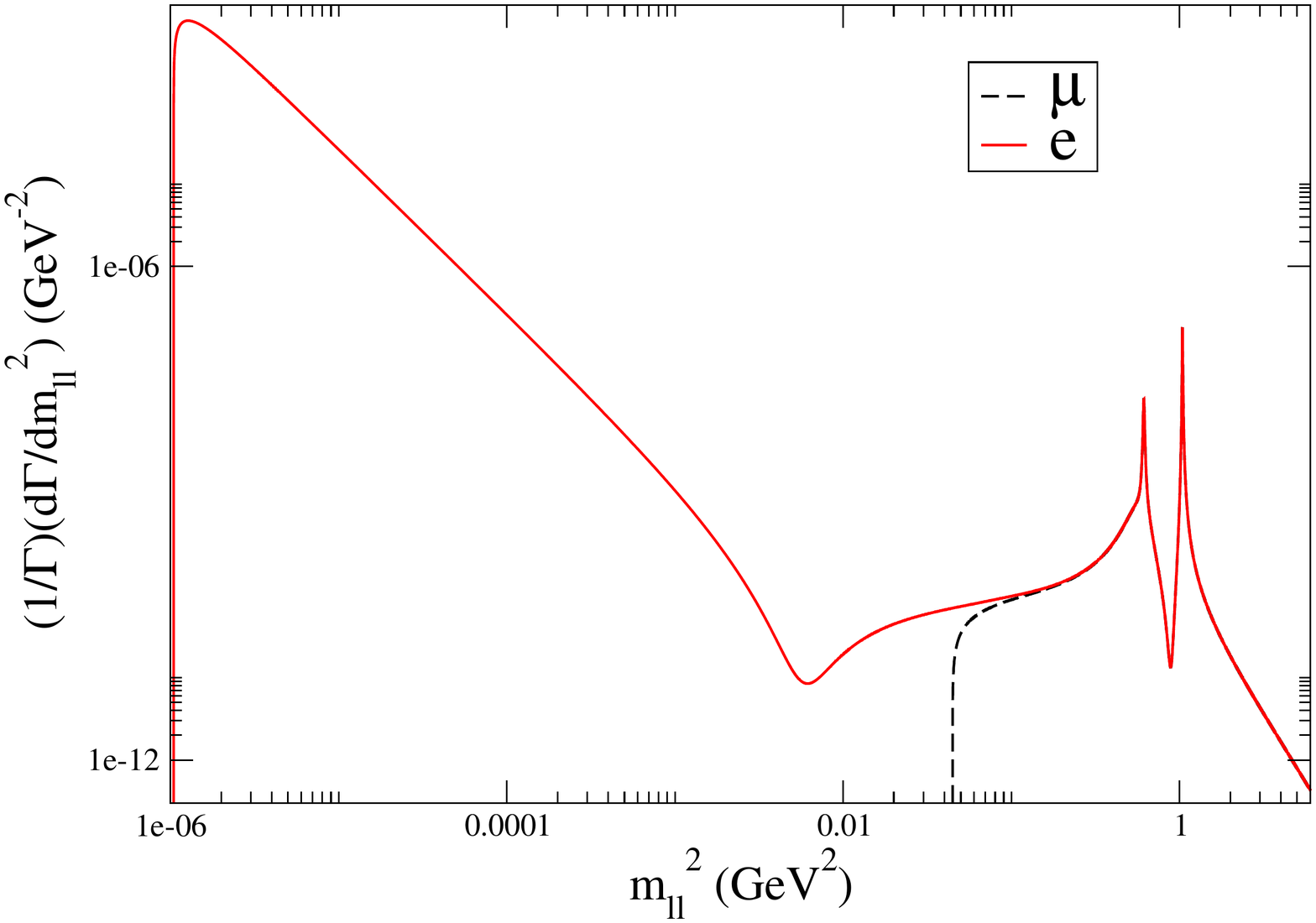}
    \caption{New LD invariant mass spectrum from kinematic threshold for $\pi$ (left) and $K$ (right). In both plots, the spectrum of {$e^+e^-$} invariant mass 
    overlaps with the $\mu^+\mu^-$ spectrum when $q^2\gtrsim0.3$ GeV$^2$; therefore, there is a negligible contribution to $R_K$.}\label{dGth}
   \end{figure}
   
 \begin{table}
  \centering
  \begin{tabular}{|c||c|c||c|}\hline
  &$P=\pi$&$P=\pi$&$P=K$\\ \hline \hline
  &$0.05 \leq q^2 \leq 8$ GeV$^2$&\ \ \ $1 \leq q^2 \leq 8$ GeV$^2$&$1\leq q^2\leq 6\text{ GeV}^2$\\ \hline
  $LD$&$(9.16\pm0.15)\cdot 10^{-9}$&$(5.47\pm0.05)\cdot 10^{-10}$&$(1.70\pm0.21)\cdot10^{-9}$\\ 
  Interf&$(-2.62\pm 0.13)\cdot 10^{-9}$& $(-2^{+2}_{-1})\cdot 10^{-10}$&$(-6\pm2)\cdot10^{-11}$\\ 
  $SD$&$(9.83^{+1.49}_{-1.04})\cdot 10^{-9}$& $(8.71^{+1.35}_{-0.90})\cdot10^{-9}$&$(1.90^{+0.69}_{-0.41})\cdot10^{-7}$\\ \hline%
 \end{tabular}\caption{LD, SD and their interference contributions to the branching ratio for both channels}\label{Tab}
 \end{table}

   The invariant mass spectrum for both channels are shown in Fig.\ref{dGth}, which shows this spectrum 
   from kinematic threshold. The difference between $\ell=e,\mu$ becomes important at $q^2\lesssim0.3$ GeV$^2$. 
   We confirm that the LHCb range is free of hadronic pollution for $P=K$, as shown in Table \ref{Tab};
   but for $P=\pi$ there is a significant pollution in the $[1,8]$ GeV$^2$ range. Comparing the 
   LHCb measurement\cite{AaijPi} to the SD 
   contribution of the branching fraction in the whole kinematic range\cite{Ali} 
   we find a better agreement by adding our LD contribution, obtaining a value 
   of $BR^{LD + SD}=(2.6^{+0.4}_{-0.3})\times10^{-8}$.
   The LD contribution induces lepton universality deviations of $\mathcal{O}(10^{-5})$ in $R_P$. The different weak 
   and strong phases of SD and LD contributions generate a CP asymmetry\cite{this,Hou} $A_{CP}=\frac{\Gamma(B^+ \to P^+ l^+ l^-) - \Gamma(B^- \to P^- l^+ 
l^-)}{\Gamma(B^+ \to P^+ l^+ l^-) + \Gamma(B^- \to P^- l^+ l^-)},$ corresponding numerical results are shown in Table \ref{Ass}.
  \begin{table}
  \centering
\begin{tabular}{|c|c|c|}\hline
    
     &$q^2\geq1$ GeV$^2$& $q^2\geq4m_\mu^2$\\ \hline
     $P=\pi$&$(2.5\pm1.5)\%$&$(14\pm2)\%$\\
     $P=K$&$-(1.3\pm0.5)\%$&$-(0.5\pm0.5)\%$\\ \hline
    
    \end{tabular}\caption{CP Asymmetry for the different energy ranges for $\pi$ and $K$.}\label{Ass}
 \end{table} 
  \section{Conclusions}
  Our analysis shows that BSM studies should be restricted to the [1,8] GeV$^2$ range for $P=\pi$; the
effect of this new LD contribution could be measured in LHCb in the next run. It is also important
to understand the current LHCb measurement of the branching fraction. In the case of $P=K$, there is 
not an important contribution for the branching fraction; this is because in the interference the peak 
of the $\phi$ resonance does not surpass the CKM suppression factor, contrary to the pure LD contribution. The $A_{CP}$ 
 we predict is an important effect that must be taken into account for BSM searches in both channels through this observable.
\section*{Acknowledgments}
We thank Cinvestav, the organizing committee and Conacyt for financial support.

\end{document}